\documentclass[12pt, preprint]{aastex}
\pdfoutput=1

\shortauthors{PULUPA AND BALE} \shorttitle{SHOCK STRUCTURE}

\begin{document}

\title{Structure on Interplanetary Shock Fronts: \\ Type II Radio
Burst Source Regions}
\author{M. Pulupa and S. D. Bale}
\affil{Physics Department and Space Sciences Laboratory, University of
California at Berkeley}
\affil{Berkeley, California, USA, 94720-7450}
\email{pulupa@ssl.berkeley.edu, bale@ssl.berkeley.edu}


\begin{abstract}

  We present \emph{in situ} observations of the source regions of
  interplanetary (IP) type II radio bursts, using data from the Wind
  spacecraft during the period 1996-2002.  We show the results of this
  survey as well as in-depth analysis of several individual events.
  Each event analyzed in detail is associated with an interplanetary
  coronal mass ejection (ICME) and an IP shock driven by the ICME.
  Immediately prior to the arrival of each shock, electron beams along
  the interplanetary magnetic field (IMF) and associated Langmuir
  waves are detected, implying magnetic connection to a
  quasiperpendicular shock front acceleration site.  These
  observations are analogous to those made in the terrestrial
  foreshock region, indicating that a similar foreshock region exists
  on IP shock fronts.  The analogy suggests that the electron
  acceleration process is a fast Fermi process, and this suggestion is
  borne out by loss cone features in the electron distribution
  functions.  The presence of a foreshock region requires nonplanar
  structure on the shock front.  Using Wind burst mode data, the
  foreshock electrons are analyzed to estimate the dimensions of the
  curved region.  We present the first measurement of the lateral,
  shock-parallel scale size of IP foreshock regions.  The presence of
  these regions on IP shock fronts can explain the fine structure
  often seen in the spectra of type II bursts.

\end{abstract}




\section{Introduction}\label{intro}

Interplanetary type II radio bursts are generated upstream of IP
shocks by solar wind electrons reflecting from the shock front.  The
reflected electron beams create Langmuir waves which produce type II
emission at the local electron plasma frequency $f_p$, and possibly
the second harmonic.  The upstream region in which the radio emission
is generated is analogous to the electron foreshock region at the
Earth's bow shock. The \emph{in situ} terrestrial foreshock has been
extensively studied by many spacecraft.  In contrast, only one
\emph{in situ} IP foreshock region has been described in the
literature. \citep{1999GeoRL..26.1573B} In this section, we will
review the basic characteristics of electron acceleration and Langmuir
wave generation at the terrestrial foreshock, and outline our
analogous measurements upstream of IP shocks.

Terrestrial foreshock electron beams were first observed in the
upstream regions of the Earth's bow shock by the ISEE
spacecraft. \citep{1979GeoRL...6..401A,1984GeoRL..11..496F} At the
terrestrial foreshock, solar wind electrons and ions are accelerated
by a fast Fermi process.  The bow shock, moving in the solar wind
frame, mirrors the particles and accelerates them tangent to the shock
along IMF lines. \citep{1984JGR....89.8857W,1984AnGeo...2..449L} The
backstreaming electrons then cause bump-on-tail velocity distributions
and generate upstream Langmuir waves.  \citep{1979JGR....84.1369F} If
the acceleration point is magnetically connected to a spacecraft, the
spacecraft observes an energetic electron beam aligned with the IMF.
The region in which these beams are present is known as the electron
foreshock region.  The Wind spacecraft has made detailed observations
of electron beams, bump-on-tail distributions, and signatures of fast
Fermi acceleration in the terrestrial
foreshock. \citep{1996GeoRL..23.1235F,1996GeoRL..23.2203L}

The efficiency of fast Fermi acceleration at curved shocks peaks when
the IMF lines are nearly tangent to the
shock. \citep{1989JGR....9415089K,1991JGR....96..143K} This places
constraints on the geometry of the shock front, as a straight upstream
IMF line has no tangent point to a shock unless curvature is present
on the shock front.  Therefore, evidence of a foreshock region is also
evidence of curved structure.  \citet{1986PASAu...6..444C} proposed a
time-of-flight mechanism for Type II emission generated by a curved IP
shock analogous to the \citet{1979JGR....84.1369F} mechanism for
emission generated by the curved terrestrial bow shock.

\citet{1998GeoRL..25.2493R} has suggested that the intermittent nature
of type II emissions implies multiple, distinct emission regions.  It
has also been shown with both remote sensing
\citep{1998JGR...10329651R,1998GeoRL..25.2493R} and \emph{in situ}
observations \citep{1999GeoRL..26.1573B} that the source region of
type II emission lies upstream of CME-driven shock fronts.  Taken
together, these observations suggest that type II emission is
generated in multiple foreshock regions upstream of IP shocks.
Theoretical models of electron reflection from the surface of
interplanetary shocks are consistent with this model, producing
electron beams and plasma radiation at $f_p$ and $2f_p$ which agree
reasonably well with the observed quantities.
\citep{2003JGRA.108c.SSH6K,2001JGR...10625041K,2003SSRv..107...27C}

We will use both `IP foreshock region' and `type II source region'
interchangeably throughout this paper, our choice of terminology
depending on whether the emphasis of the discussion is on the
accelerated electrons or the radio emission.  Both terms refer to the
same physical region.

The event described by \citet{1999GeoRL..26.1573B} was the first
observed \emph{in situ} measurement of a type II radio burst.  We have
examined the data set of IP shocks observed by the Wind spacecraft,
searching for additional events.  Section \ref{eventsel} describes the
results of the search.

Section \ref{observ} presents detailed \emph{in situ} observations of
three selected IP foreshock regions, showing the correlation between
upstream electron beams and the local generation of type II radiation.

\begin{figure}
  \figurenum{1} \epsscale{1.0}
  \includegraphics[angle=270]{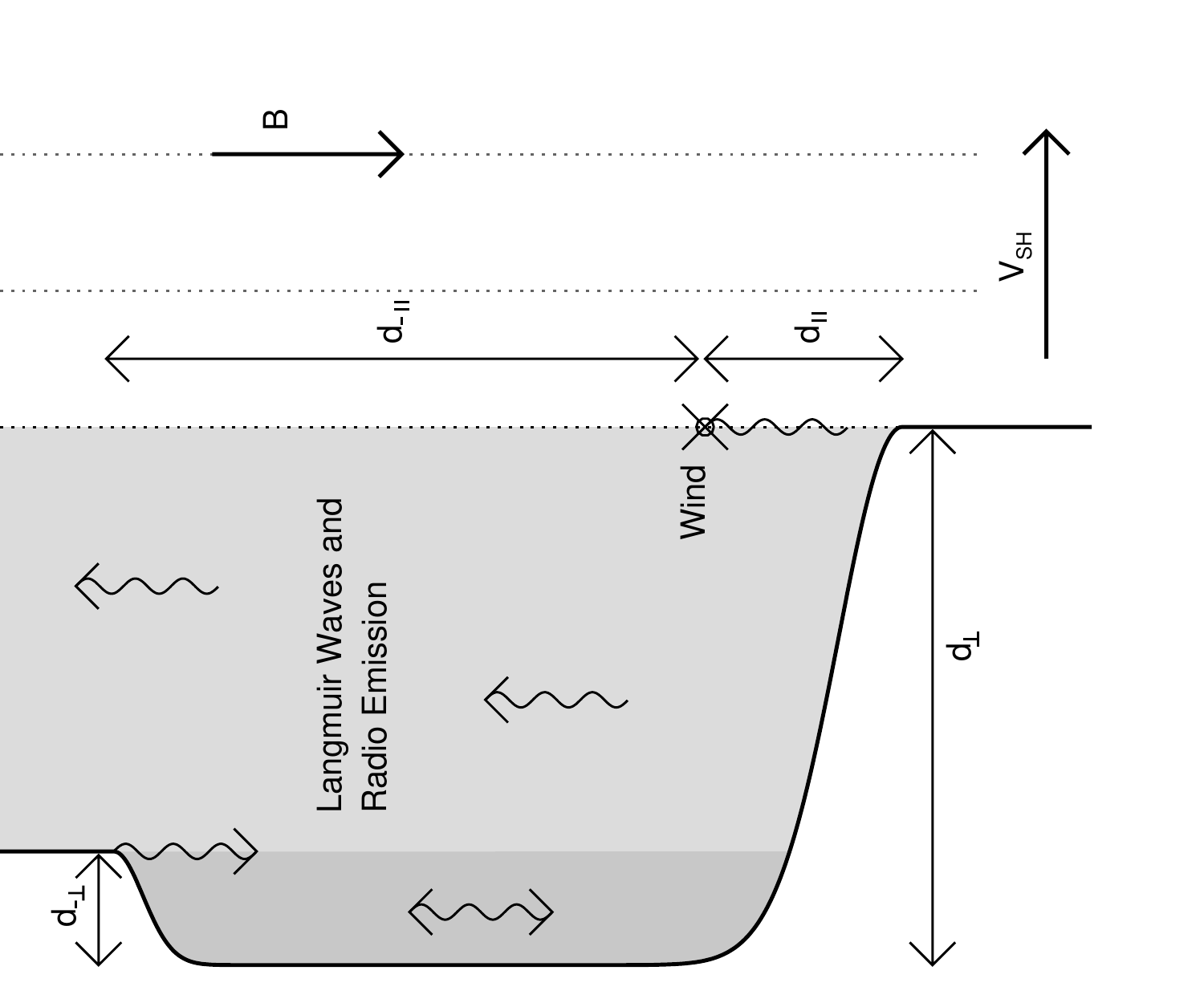}
  \caption{A cartoon of shock structure consistent with our
  observations.  Electron flux in the $\mathbf{B}$ and $-\mathbf{B}$
  direction increases previous to shock arrival, as Wind is connected
  along the IMF line to an advanced section of the shock front.
  Langmuir waves and electron beams are observed in the foreshock
  region. \label{cartoon}}
\end{figure}

Sections \ref{height} and \ref{distance} emphasize the information
about shock structure that may be deduced from the velocity-dispersed
foreshock electron beams.  By analyzing velocity-dispersed electron
beams in the foreshock region, we can determine the shock-parallel and
perpendicular scale size of the shock front structure.  The calculated
parameters are illustrated in Figure \ref{cartoon}.  In order to
calculate the perpendicular scale height $d_\perp$ of the shock
structure, we must determine both the shock speed $V_{sh}$ and the
initial acceleration time of the foreshock electrons $t_{0}$.  To
calculate the lateral distance $d_\parallel$ from the spacecraft to
the acceleration point, we analyze the velocity dispersion of the
foreshock electron beam.  Since the spacecraft can be connected to the
shock front in both the IMF-parallel and antiparallel direction, we
can potentially determine $d_{-\perp}$ and $d_{-\parallel}$ as well.
Taken together, these measurements describe the nature of the rippling
which occurs along the shock front.  We present all distances in units
of $R_E$ as well as $\mathrm{km}$, to facilitate comparison with the
terrestrial foreshock region.  The shock surface shown in Figure
\ref{cartoon} is approximately to scale with the $d_{\pm \perp}$ and
$d_{\pm \parallel}$ parameters determined for the 28 August 1998
shock.

Section \ref{coplanarity} examines the validity of the assumptions we
use in analyzing the IP shocks, and Section \ref{discussion}
summarizes our observations and discusses possible origins of the IP
foreshock regions.

\section{Event Selection}\label{eventsel}

We have investigated \emph{in situ} data from the Wind spacecraft for
several hundred shocks which occurred during the time period
1996-2002.  Our list of shocks was obtained from the MIT database of
Wind shock
crossings.\footnote{http://space.mit.edu/home/jck/shockdb/shockdb.html}
We used data from the Wind/WAVES plasma wave experiment
\citep{1995SSRv...71..231B} to investigate each shock.  Of the 377
shock crossings in the database, we found 125 events which upon visual
inspection contained possible foreshock Langmuir wave activity (LWA),
as evinced by strong plasma frequency radiation immediately prior to
shock arrival.  We inspected these events closely for signs of
\emph{in situ} type II radiation.

We eliminated events with rapid changes in plasma density and magnetic
field prior to shock arrival, in order to avoid misidentification of
upstream waves as foreshock structures.  We also eliminated events
with other possible sources of plasma frequency emission, such as
Langmuir waves caused by reflection from the terrestrial foreshock, or
type III radio bursts arriving at Earth.

Using data from the Three-Dimensional Plasma (3DP) instrument suite on
Wind \citep{1995SSRv...71..125L}, we searched for correlations between
IMF-parallel electron beams and LWA.  The electron beams were measured
by the low energy electron electrostatic analyzer (EESA-L), an
instrument on the 3DP suite.  The EESA-L instrument measures one full 3D
electron distribution function per spacecraft spin (3
seconds). However, the cadence of data in the telemetry stream is
determined by the telemetry rate of the spacecraft. In normal
operation, the spacecraft returns distribution functions at a rate of
approximately one per 100 seconds.  An instrument `burst-mode'
provides full time resolution (3 second) measurements when a
burst-mode trigger criterion is met.  The trigger is computed on board
from a selectable set of measurements (e.g. ion or electron flux
changes). Some programmed burst triggers are optimized to catch
shocks, while others might be optimized to investigate energetic
particle events.  When an event is detected by the burst mode trigger,
the spacecraft stores higher cadence data into a circular buffer, and
sends the data when the event ends or when the memory is full.

In many cases, the LWA occurred in short bursts lasting less than
1 minute, and therefore did not appear in the Wind low cadence
data. Wind was operating in burst mode for less than half of the
shocks with possible Langmuir wave activity.  On one occasion, the LWA
was sustained for minutes prior to the shock, and could therefore be
correlated with the low cadence electron data.  The majority of the
125 events with possible \emph{in situ} type II radiation were
eliminated from consideration because the lack of burst mode data
made the association between the plasma emission and IMF-parallel
electron beams impossible to confirm.

In order to determine the source of the IP shocks, we used the lists
of CME events and related shocks published in
\citet{2003JGRA.108d.SSH6C} and \citet{2004JGRA..10906109M}, as well
as the type II radio burst list maintained at Goddard Space Flight
Center.\footnote{http://lep694.gsfc.nasa.gov/waves/waves.html}

We found a total of 8 events (including the event published in
\citet{1999GeoRL..26.1573B}) possessing all of the characteristics
described above: upstream LWA, observed IMF-parallel electron beams
correlated with the LWA, a relatively stable upstream plasma
environment during the periods of LWA, and an identifiable ICME source
for the shock.  These events are listed in Table \ref{event_tab}.

Of the 8 events, three contained velocity-dispersed electron beams.
As will be shown in the following sections, this feature enables the
measurement of the lateral scale size of the shock front structures
where type II radiation is generated.  The measurement of this lateral
scale size is the primary new measurement presented in this paper,
therefore we will focus on the three events with velocity-dispersed
beams.

\section{Foreshock Electron Observations}\label{observ}

The three \emph{in situ} type II events with observed velocity
dispersed electron beams occurred upstream of IP shocks which arrived
at the Wind spacecraft on 15 May 1997, 26 August 1998, and 11 February
2000.  The 26 August 1998 event has been described in
\citet{1999GeoRL..26.1573B}.  The shock for that event was driven by
an ICME associated with an X1.0 class flare which occurred at 22:09 UT
on 24 August 1998.

The 15 May 1997 (11 February 2000) shock was driven by an ICME
associated with a C1.3 (C7.3) class flare which occurred at 04:55 UT
on 12 May 1997 (02:08 UT on 10 February 2000.)

\begin{figure}
  \figurenum{2} \epsscale{1.0} \plotone{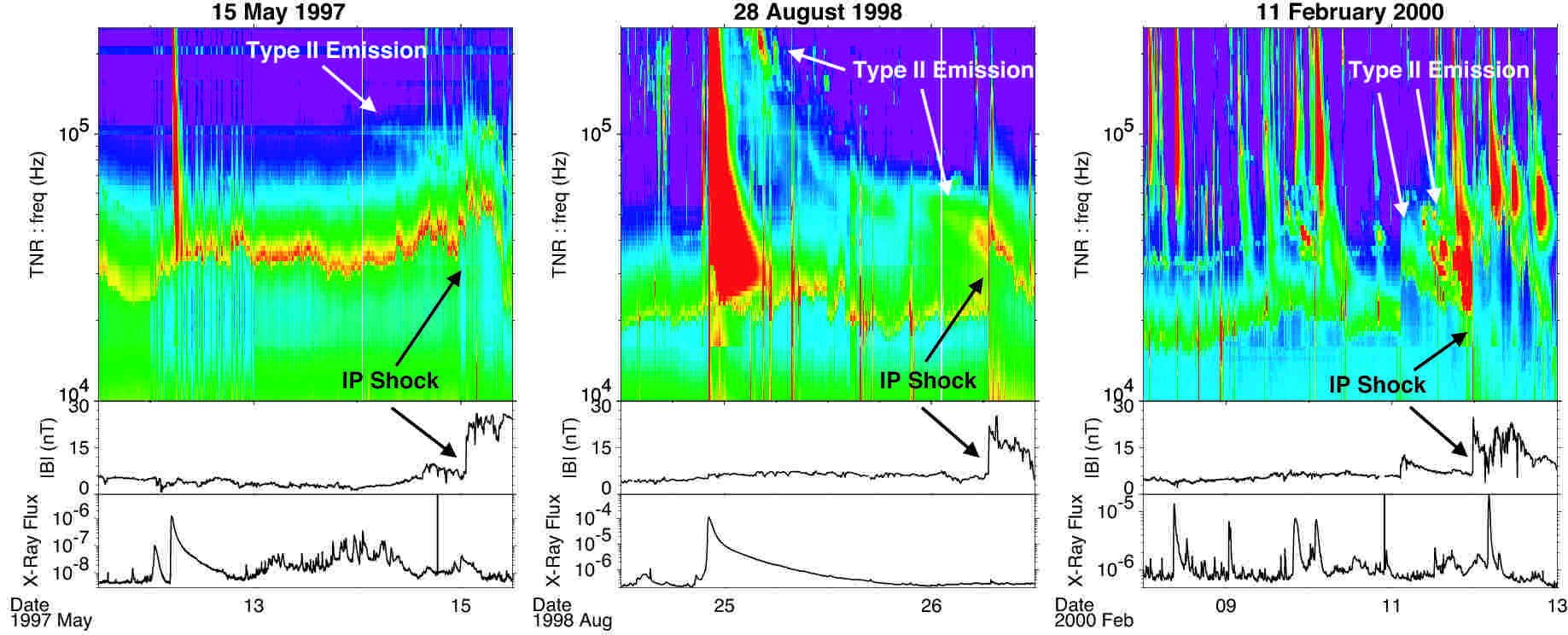}
  \caption{Radio wave, magnetic field, and GOES x-ray data for three
  shock crossings seen by the Wind spacecraft.  The top panel is a
  dynamic spectrum from the WAVES instrument on Wind, the bottom panel
  is the x-ray flux data in the band $1-8 \mbox{\AA}$ from the GOES-8
  satellite (for May 1997) and the GOES-10 satellite (for August 1998
  and February 2000.)  The flare activity is shown by the x-ray peaks
  and type III radio bursts on 12 May 1997, 24 August 1998, and 9
  February 2000.  The type II emissions can be seen as slowly drifting
  features in the spectrum, and the spacecraft shock crossings are
  indicated by abrupt jumps in the local plasma frequency and the
  magnetic field. \label{wavfig}}
\end{figure}

Figure \ref{wavfig} shows dynamic spectra from Wind/WAVES and magnetic
field data from the MFI instrument \citep{1995SSRv...71..207L} on
Wind, along with GOES X-ray data for each of these three events.
Upstream and downstream plasma parameters for each shock are listed in
Table \ref{plasma_tab}.  At each event, the Wind spacecraft was in the
foreshock region for a timespan of 20 to 40 seconds.

\begin{figure}
  \figurenum{3} \epsscale{0.8} \plotone{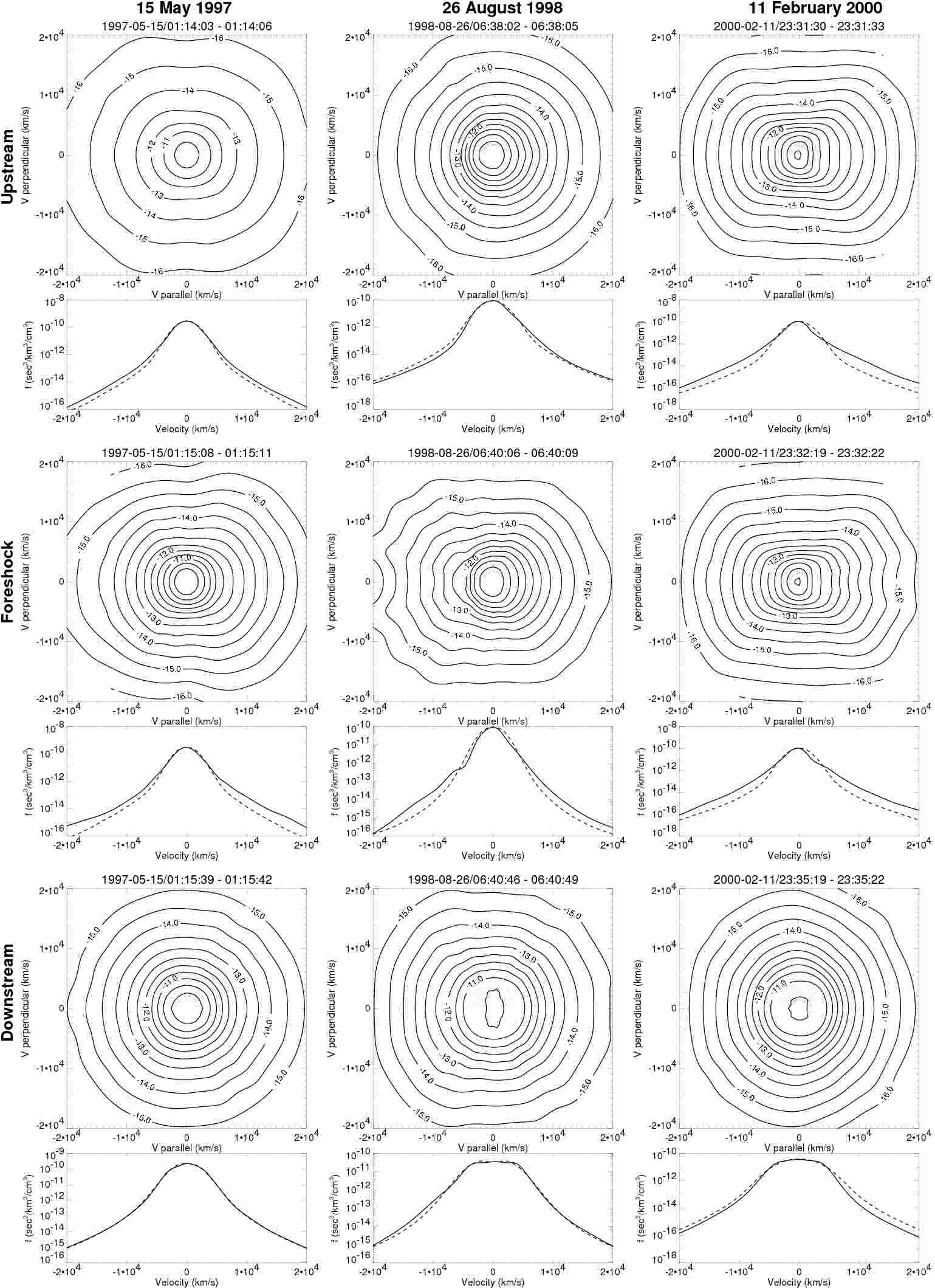}
  \caption{Electron velocity distributions measured by the EESA-L
  instrument on Wind during the upstream pre-foreshock, foreshock, and
  downstream periods for the three IP shocks.  The foreshock region is
  characterized by bumps on the parallel distribution function and the
  loss cone evident in the 26 August 1998 and 11 February 2000
  foreshock regions.  The bottom part of each panel shows a parallel
  (solid line) and perpendicular (dashed line) cut through each
  two-dimensional distribution. \label{distfig}}
\end{figure}

Figure \ref{distfig} shows two-dimensional electron pitch angle
distributions from the EESA-L instrument on Wind/3DP.  The
distributions are shown for each event in the upstream `pre-foreshock'
region, in the foreshock region, and in the downstream region after
the shock has passed.  The foreshock region is distinguished by the
electron beams in the IMF-parallel direction, which can be seen as the
bulges in the parallel and anti-parallel direction on the
two-dimensional distributions and in the parallel (solid line) and
perpendicular (dashed line) cuts.  The observed electron distribution
functions are consistent with the predictions of electron beams
originating from the shock as predicted by
\citet{1979JGR....84.1369F}, and reflected by the Fast Fermi process
described by \citet{1984JGR....89.8857W} and
\citet{1984AnGeo...2..449L}.  The distribution functions also show an
angular feature corresponding to a loss cone.  This loss cone feature
is predicted by the Fast Fermi theory and has been observed by
Wind/3DP in the terrestrial foreshock.  \citep{1996GeoRL..23.2203L}

The electron beam reflected from the surface of the shock creates a
bump on the tail of the electron distribution function.  Due to
velocity selection effects, this bump is most prominent at the
boundary of the foreshock region.  \citep{1984GeoRL..11..496F} The
Wind Solar Wind Experiment (SWE) \citep{1995SSRv...71...55O} has
observed the positive slope at many encounters with the terrestrial
foreshock boundary. \citep{1996GeoRL..23.1235F} However, the Wind/3DP
instrument has insufficient energy resolution to resolve the
positively sloped region on the tail of the distribution function, and
therefore does not observe the bump during the same encounters.

After the arrival of the shock, the distributions display the
broadened, flat-topped characteristics common to distributions
downstream of strong interplanetary shocks.
\citep{2003JGRA.108l.SSH1F}

\begin{figure}
  \figurenum{4} \epsscale{1.0} \plotone{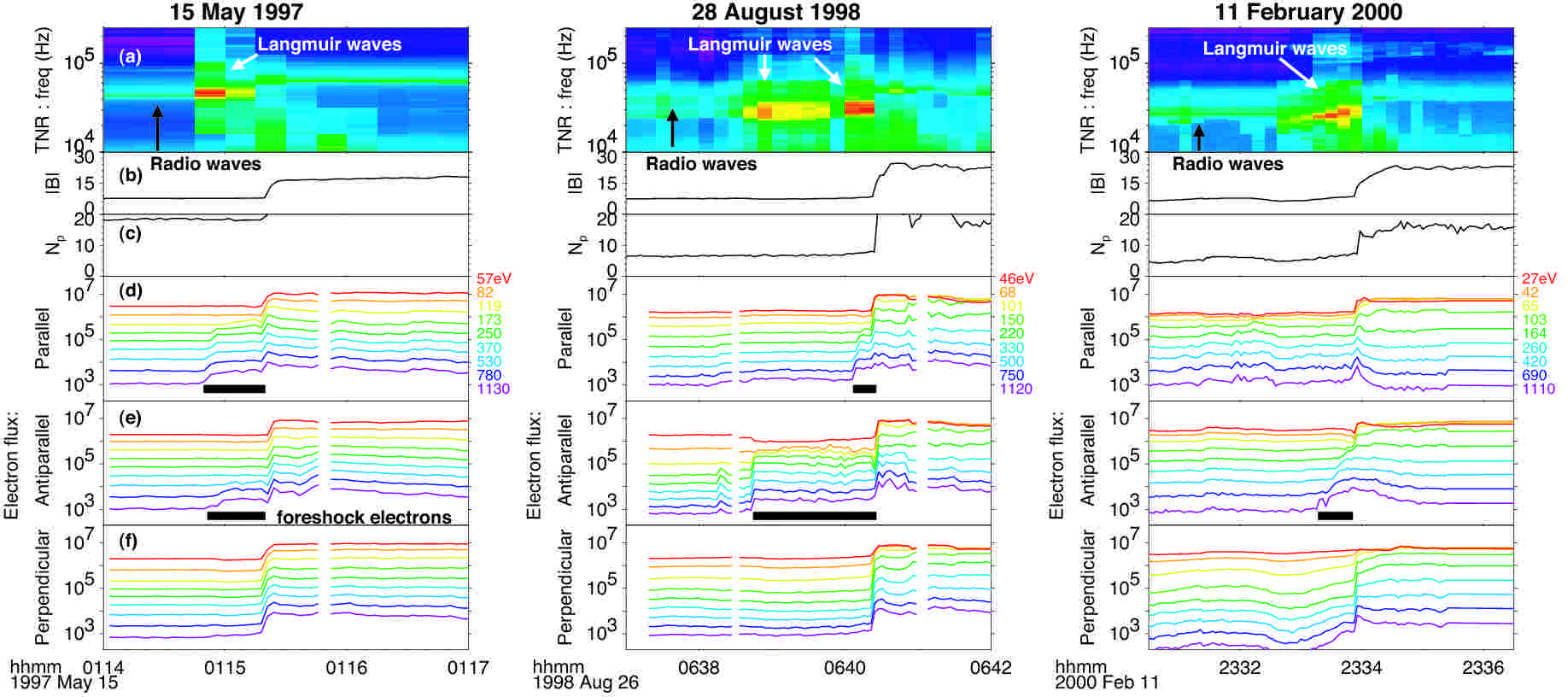}
  \caption{\emph{In situ} particle and wave data from the Wind
  spacecraft for the three shock crossings.  Panel (a) is a dynamic
  spectrum from the WAVES Thermal Noise Receiver, showing Langmuir
  wave activity in the foreshock region.  Panels (b) and (c) are
  magnetic field and density measurements from MFI and 3DP, both
  showing a jump at the arrival of the shock.  Panels (d), (e), and
  (f) are electron flux energy distributions from the \mbox{EESA-L}
  experiment on 3DP, in the parallel, antiparallel, and perpendicular
  directions. The foreshock electron beams are denoted by black bars
  in the parallel and antiparallel panels.  The units for magnetic
  field are $\rm nT$, for density $\rm 1/cm^3$, and electron flux
  $\rm 1/eV/sec/cm^2/ster$. \label{langfig}}
\end{figure}

The association of the foreshock electrons with the type II emission
is established using the Langmuir wave observations from the WAVES
instrument.  Figure \ref{langfig} shows wave and particle data from
Wind at each shock crossing.  Panel (a) is a WAVES dynamic spectrum
showing intense Langmuir wave activity in each foreshock region.
Panels (b) and (c) show magnetic field magnitude from MFI and proton
density from 3DP.  In both panels, there is a clear discontinuity as
each shock crosses the spacecraft.  The bottom three panels show
electron energy flux for a range of energies measured by the low
geometric factor Electron Electrostatic Analyzer (EESA-L) on 3DP.  The
Wind spacecraft was in burst mode during each shock crossing,
measuring full three dimensional electron distributions once every
three seconds.  Panels (d), (e), and (f) show the IMF-parallel,
antiparallel, and perpendicular fluxes, respectively.  The Langmuir
waves in panel (a) are associated with increases in electron flux in
both the parallel and antiparallel directions, except for the 11
February 2000 shock, for which only antiparallel foreshock flux was
observed.  The black bars on the plots indicate the locations of
elevated flux due to foreshock electrons.  The correlation between
foreshock electrons and Langmuir wave activity strongly indicates that
these regions are sources of type II radio emission.  In the following
sections, we will use the electron burst data to characterize the
shock-perpendicular scale height and shock-parallel scale distance of
the foreshock region.

\section{Shock-Perpendicular Scale Height}\label{height}

The scale height $d_\perp$ of the feature on the shock front is
determined by the speed of the shock in the spacecraft frame
$V_{sh}^{S/C}$ and the amount of time $\Delta t$ between the start of
foreshock electron enhancement and arrival of the shock.
\begin{equation}
  d_\perp \approx V_{sh}^{S/C} \cdot \Delta t
\end{equation}

The time interval, indicated by the black bars in Figure
\ref{langfig}, is easily measured.  The shock velocity in the
spacecraft frame is determined by mass flux conservation across the
shock boundary \citep{2000ESASP.449...99P}, and is given by:
\begin{equation}
  V_{sh}^{S/C} = \frac{\Delta(\rho\mathbf{V}^{S/C})}{\Delta{\rho}} \cdot \hat{\mathbf{n}}
\end{equation}
where $\rho$ is the local mass density and $\hat\mathbf{n}$ is the
unit vector normal to the shock surface.  The determination of
$\hat\mathbf{n}$ is discussed in a later section.

The scale height of the 26 August 1998 shock was calculated in
\citet{1999GeoRL..26.1573B} and found to be
$136,000\: \mathrm{km}\:(21.3\:R_E)$ for $d_{-\perp}$, the height of
the structure in the antiparallel direction, and
$25,000\:\mathrm{km}\:(3.9\:R_E)$ for $d_{\perp}$, in the parallel
direction.  Our method yields the significantly smaller values of
$69,000\: \mathrm{km}\:(10.1\:R_E)$ for $d_{-\perp}$ and
$15,000\: \mathrm{km}\:(3.9\:R_E)$ for $d_{\perp}$.  These results
differ because \citet{1999GeoRL..26.1573B} uses the total time between
the start of the flare and arrival of the shock to calculate an
average shock speed from the inner heliosphere to 1 AU.  Here we use
the \emph{in situ} method described above, which yields an
instantaneous $V_{sh}$ that more accurately describes the local shock
parameters.

The calculated values of $d_{\perp}$ and $d_{-\perp}$ for each of the
three analyzed shocks are listed in Table \ref{struct_tab}.

\section{Estimating Shock-Parallel Distance}\label{distance}

When electrons reflect from the shock surface and stream along IMF
lines to the spacecraft, the most energetic accelerated electrons will
arrive first, followed by the lower energy electrons.  Provided that
the distance from the acceleration point is sufficiently large and the
energy and time resolution of the detector is sufficiently good, this
time of flight dispersion is observable in the foreshock electron
beam.

The velocity of the electrons is determined by the (nonrelativistic)
formula
\begin{equation}
  v_e=\sqrt{2E/m_e}
\end{equation}
where E is the kinetic energy of the electron.

We assume that the electrons were accelerated instantaneously at a
time $t_{0}$.  The transit time for each energy bin is determined by
the time interval between $t_{0}$ and $t_{onset}$, when the first
enhancement appeared in that bin.  The parallel distance $d_\parallel$
and the initial acceleration time $t_0$ are determined by fitting the
measured values of $v_e$ and $t_{onset}$ to the simple functional form
\begin{equation}\label{d_par_eq_1}
  d_\parallel=v_e(t_{onset}-t_{0})=v_e \cdot \Delta t
\end{equation}

\begin{figure}
  \figurenum{5} \epsscale{1.0} \plotone{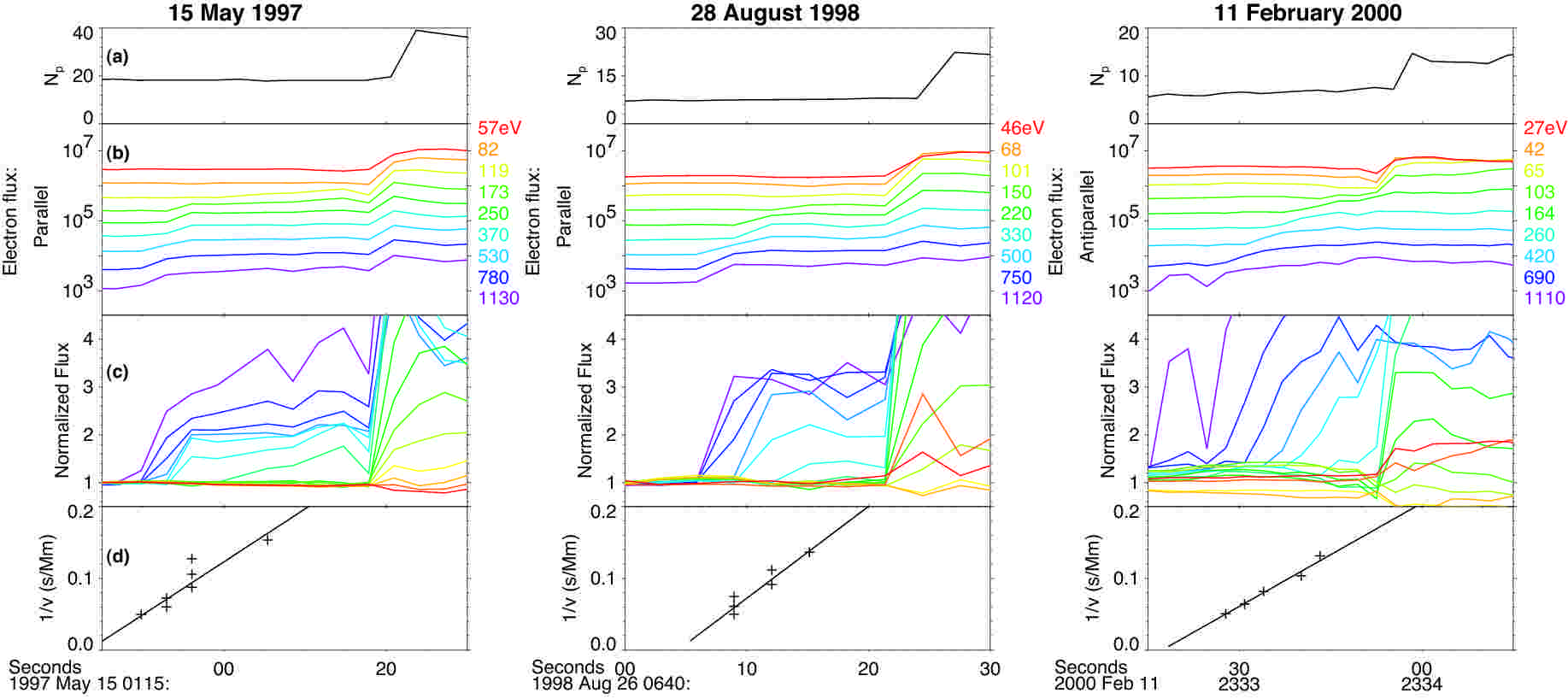}
  \caption{\emph{In situ} particle data from the Wind 3DP instrument.
  Panel (a) is the particle density, showing the shock arrival time.
  Panel (b) is the parallel (antiparallel for 11 February 2000)
  electron energy distribution, showing the velocity dispersed
  electron beam.  Panel (c) emphasizes the velocity dispersion by
  normalizing each energy channel to its pre-foreshock flux. Panel (d)
  shows a fit of arrival time to inverse electron velocity, as
  described in Equation \ref{d_par_eq_1}. \label{elbfig}}
\end{figure}

The results of this fit are shown in Figure \ref{elbfig}, which
contains \emph{in situ} data from the Wind/3DP in the foreshock
region, for each of the three analyzed shocks.  For each shock, only
one of two possible directions contained sufficient velocity
dispersion in the electron beam that the above equation could be fit.
For the 15 May 1997 and 28 August 1998 shock, the parallel direction
was fit.  For the 11 February 2000 shock, the antiparallel direction
was fit.

The proton density is plotted in panel (a).  The jump in density
indicates arrival of the shock.  Panel (b) shows the flux of electrons
parallel (or antiparallel for 11 February 2000) to $\mathbf{B}$ prior
to shock arrival.  Note that flux enhancement occurs first in the high
energy electron bins.  Panel (c) shows the same electron flux, with
each energy bin normalized to its preshock level.  The onset time for
each energy bin is defined as the time when the normalized flux first
rises past a threshold value.  Panel (d) shows a fit of onset time
against inverse velocity for each energy bin which showed foreshock
enhancement.  $t_{acc}$ and $d_\parallel$ are given by fitting the
velocity and time data to Equation \ref{d_par_eq_1}.

It is important to clarify the relationship between Equation
\ref{d_par_eq_1} and the `foreshock coordinate system' established by
\citet{1979JGR....84.1369F}.  \citet{1979JGR....84.1369F} noted that
velocity-dispersed electrons will be a steady-state spatial feature of
an electron foreshock in the shock frame.  A `cutoff' velocity $v_c$
exists below which the shock-accelerated electrons will not reach the
spacecraft; the result is a beam-like feature that is unstable to
Langmuir waves.  \citet{1979JGR....84.1369F} showed that
$v_c \approx v_{sw} d_\parallel /DIFF$, where $DIFF$ is the distance
downstream from the first tangent field line to the shock, in the
direction of the solar wind flow.  In our formulation,
$DIFF = v_{sh} \Delta t \approx v_{sw} \Delta t$ so that
$d_\parallel = v_c \Delta t$, which is equivalent to Equation
\ref{d_par_eq_1}.  Hence this is just a transformation from the shock
frame to the spacecraft frame.

For the 26 August 1998 shock discussed in \citet{1999GeoRL..26.1573B},
the fit value for the shock parallel distance $d_{\parallel}$ is
$78,000\: \mathrm{km}\:(12.2\:R_E)$.  The parallel beam for the 15 May
1997 shock and the antiparallel beam for the 11 February 2000 shock
were also fit using this method, yielding
$d_\parallel = 136,000\: \mathrm{km}\:(21.2\:R_E)$ for the 15 May 1997
shock and $d_{-\parallel} = 151,000\: \mathrm{km}\:(32.6\:R_E)$ for
the 11 February 2000 shock.

The antiparallel foreshock beams seen in the 15 May 1997 and 26 August
1998 shocks do not display velocity dispersed onset times, implying
that the acceleration site was close to the spacecraft.  An upper
limit on $d_{-\parallel}$ may be obtained by noting that if the
fastest and slowest foreshock electrons arrived at the spacecraft at
the same time to within the time resolution of 3DP burst mode, then
$d_{-\parallel}$ must satisfy
\begin{equation}\label{d_par_eq_2}
  d_{-\parallel} /v_{slow} - d_{-\parallel}/v_{fast} \leq 3~\mathrm{seconds}
\end{equation}
which yields an upper limit for
$d_{-\parallel} \leq 140,000\: \mathrm{km}\:(21.9\:R_E)$ for 15 May
1997 and $d_{-\parallel} \leq 26,000\: \mathrm{km}\:(4.0\:R_E)$ for 28
August 1998.

The calculated values of $d_{\parallel}$ and $d_{-\parallel}$ for all
of the analyzed shocks are listed in Table \ref{struct_tab}.

\section{Upstream IMF and coplanarity of shock
front}\label{coplanarity}

In our calculation of $d_\parallel$ and $d_\perp$, we have made two
assumptions about the magnetic field: that the IMF line connecting the
shock to the spacecraft is straight, and that the shock propagation
direction is perpendicular to the IMF lines.  In this section, we
investigate the validity of these assumptions.

The boundary of the electron foreshock region is determined by the IMF
line tangent to the shock surface.  Turbulence in the solar wind and
electromagnetic radiation generated by shock-accelerated particles can
deform the structure of the IMF.  Numerical models of magnetic field
line transport have been developed to simulate IMF conditions at
planetary bow shocks.  \citep{1996GeoRL..23..793Z}

For IMF conditions similar to those at 1 AU, the ratio of the spread
in the foreshock boundary $\Delta r$ to the length of the connecting
IMF line $r$ is $\Delta r/r \approx 0.1$.  If
$\Delta r/r > d_{\perp}/d_{\parallel}$, then the foreshock
measurements could be explained simply as an effect of turbulence in
the IMF.  However, for each analyzed shock, $d_{\perp}/d_{\parallel}$
(and $d_{-\perp}/d_{-\parallel}$) is greater than $0.1$, so magnetic
turbulence alone cannot account for the apparent structure on the
shock front.

If the IMF lines are straight, but the shock front is not coplanar
with the IMF lines, then the time of flight dispersion analysis can
yield misleading results for the perpendicular distance to the shock.
The above analysis assumes a coplanar structure, so we must establish
that this is a good approximation.

The orientation of the shock to the IMF can be determined by mixed
mode (including both field and particle data) coplanarity analysis.
The coplanarity theorem for compressive shocks states that the shock
normal ($\mathbf{\hat{n}}$), the upstream and downstream magnetic
fields ($\mathbf{B}_u$ and $\mathbf{B}_d$), and the velocity jump
across the shock ($\Delta \mathbf{ V}$) all lie in the same plane.  If
$\Delta \mathbf{B} \equiv \mathbf{B}_d-\mathbf{B}_u$ is the change in
magnetic field, then the shock normal \citep{2000ESASP.449...99P} is
given by:
\begin{equation}\label{eqcop}
  {\mathbf{\hat{n}} =
  \frac{(\Delta \mathbf{B} \times \Delta \mathbf{V}) \times \Delta
  \mathbf{B}}{|(\Delta \mathbf{B} \times \Delta \mathbf{V}) \times
  \Delta \mathbf{B}|}}
\end{equation}

The perpendicularity of the shock is measured by $\theta_{bn}$, the
angle between the upstream magnetic field $\mathbf{B}_u$ and
$\mathbf{\hat{n}}$.  To check consistency, we also calculate
$\theta_{bn}$ using $\mathbf{B}_u$ and $\mathbf{B}_d$ in place of
$\Delta \mathbf{B}$ in Equation \ref{eqcop}.
\citep{2000ESASP.449...99P} Values for $\theta_{bn}$ for each shock
are listed in Table \ref{plasma_tab}.

For each mixed mode calculation at each shock,
$\theta_{bn} > 80^{\circ}$, so the assumption of a locally
perpendicular shock front at Wind is a good one, and the angle between
the shock front and magnetic field does not introduce large errors in
our calculation of $d_{\parallel}$ or $d_{\perp}$.

\section{Discussion}\label{discussion}

We have conducted a survey of several hundred IP shocks, and found
\emph{in situ} type II radiation, correlated with IMF-parallel
electron beams, present at eight IP shocks.  Most IP shocks do not
show evidence of \emph{in situ} type II radiation, and of those that
do, few show evidence of upstream electron beams observable by the
Wind spacecraft.  However, this does not disprove the IP foreshock
mechanism as the generator of type II radiation.  The proposed
mechanism is a localized phenomenon, while the consequent radiation is
visible throughout the heliosphere.  It is quite unlikely that any
given region of localized emission will be encountered by the Wind
spacecraft.  The exact probability of such an encounter depends on the
size and number of IP foreshock regions, and this paper represents a
first attempt at quantifying both.  Of the events which do present
observed \emph{in situ} type II radiation, the low number of events
with correlated waves and electron beams is primarily due to the
infrequent availability of high cadence measurements of the electron
distributions.

In addition to the survey of the Wind data set, we present detailed
\emph{in situ} observations of the electron foreshock region of three
IP shocks.  In each of these three events, the presence of
velocity dispersion in the foreshock electron measurements allows
calculation of the parallel and perpendicular scale size of shock
front structure.  The 15 May 1997 and 28 August 1998 shocks have
evidence for foreshock structure in both the parallel and antiparallel
directions, suggesting the presence of a bay in which electron beams
can be mirrored and accelerated, generating Langmuir waves and radio
emission.

Although foreshock regions upstream of IP shocks imply curved
structures, the foreshock regions could theoretically be created by
either a curved magnetic field or by a curved shock.  Using only
measurements made by a single spacecraft, it is impossible to
determine which effect predominates. Previous studies focused on ion
acceleration have assumed both cases: propagation of a planar shock
through a region of curved magnetic fields
\citep{1994ApJS...90..553E}, or ion acceleration by repeated
encounters with a rippled shock. \citep{1990JGR....9511993D} It is
shown in the previous section that upstream magnetic turbulence alone
cannot explain the dimensions of the acceleration regions, and
therefore at least a portion of the foreshock region must be created
by shock front structure.  Regardless of which effect predominates,
the methodology used in this paper to estimate the characteristic
dimensions of the foreshock regions is valid.

It is unclear at present what causes the observed shock front
structure.  The curvature may be caused by Alfv\'en speed
inhomogeneities in the solar wind, which can allow different sections
of the shock to propagate at different speeds through the heliosphere.
Shock reformation, a process in which protons reflected from a shock
surface generate upstream instabilities which lead to formation of a
new shock front upstream of the original front, may also play a role.
One-dimensional hybrid simulations suggest that shock reformation in
perpendicular shocks depends on upstream parameters such as Mach
number and plasma $\beta$.  \citep{2002GeoRL..29x..87H} However, more
recent two-dimensional studies suggest that perpendicular shock fronts
may be dominated by whistler waves, which can inhibit reformation.
\citep{2007GeoRL..3414109H}

Multi-spacecraft missions such as STEREO will be greatly helpful in
future investigations of shock structure, and future studies with
multi-point measurements should improve current estimates of the
frequency and size of IP foreshock regions, which will provide useful
input for models of type II generation.  If shock front structure
is a common feature of IP shocks, then each foreshock region on a
shock front would create an independent source of type II radio
emission.  The spatial variation in upstream plasma density at these
multiple source regions could then be responsible for the fine
structure observed in many type II bursts.


\acknowledgments
  The authors would like to thank J. C. Kasper for use of the Wind IP
  shock list.  Work at UC Berkeley is sponsored by NASA grants
  NNG05GH18G and NNX06AF25G to the University of California.  MPP is
  supported by the NASA/GSRP grant NNG04GN52H.  Wind/MFI data is
  courtesy of the MFI team (PI: R. P. Lepping) at Goddard Space Flight
  Center.  GOES x-ray data is obtained from the National Geophysical
  Data Center at NOAA.


\bibliographystyle{apj}
\bibliography{/Users/pulupa/Documents/latex/pulupa_bib}

\clearpage


\begin{deluxetable}{rrr|rrr|ccc}
  \tabletypesize{\small} \tablenum{1}
  \tablecaption{IP Shocks With Observed \emph{in situ} Type II Source
  Regions \label{event_tab}} 
\startdata
  \multicolumn{3}{c|}{LASCO CME}& \multicolumn{3}{c|}{IP Shock} &
  Burst & Velocity & Type II
  \\
  Year & Date & Time & Year & Date & Time & Data & Dispersion &
  Emission \tablenotemark{a}
  \\
  1997 & May 12 & 0530 & 1997 & May 15 & 0115 & X & X & X
  \\
  1998 & Aug 24 & 2209 & 1998 & Aug 26 & 0640 & X & X & X
  \\
  2000 & Feb 10 & 0230 & 2000 & Feb 11 & 2333 & X & X & X
  \\
  2000 & Feb 17 & 0431 & 2000 & Feb 20 & 2045 & X & & X
  \\
  2000 & Oct 02 & 0350 & 2000 & Oct 05 & 0240 & X & &
  \\
  2000 & Oct 09 & 2350 & 2000 & Oct 12 & 2145 & X & &
  \\
  2001 & Mar 19 & 0526 & 2001 & Mar 22 & 1355 & & &
  \\
  2001 & Dec 26 & 0530 & 2001 & Dec 30 & 2005 & X & & X
  \\
  \enddata
  \tablenotetext{a}{Denotes presence in the Wind/WAVES type II
  database.  \emph{In situ} plasma frequency (type II) radiation was
  observed in the foreshock region of each event.}

\end{deluxetable}

\begin{deluxetable}{lrrrrrrrr}
  \tabletypesize{\small} \tablenum{2}
  \tablecaption{Shock and Plasma Parameters for Selected
  Events \label{plasma_tab}} 
\startdata Date &$\mathrm{UT}$
  &$\mathrm{B_1/B_0}$ &$V_{sh}^{S/C}\:(\mathrm{km/s})$
  &$\mathrm{U_{sw}\:(km/s)}$ &$\mathrm{M_A}$ &$\beta$
  &$r_{L}\:(\mathrm{km})$
  &$\theta_{bn}$\\
  15 May 1997     	&0115&2.19&423.&321.&2.62&0.44&392.&89.9\\
  26 August 1998	&0640&3.09&659.&493.&2.63&0.66&665.&88.1\\
  11 February 2000	&2333&3.24&678.&458.&3.19&0.39&654.&85.7\\
  \enddata

\end{deluxetable}

\begin{deluxetable}{lrrrrrrrrrr}
  \tabletypesize{\small} \tablenum{3}
  \tablecaption{Shock Structure Parameters ($d$ measured in Mm
  ($R_E$)) \label{struct_tab}}
\startdata
  Date&\multicolumn{1}{c}{$t_0$}&\multicolumn{1}{c}{$t_{shock}$}
  &\multicolumn{2}{c}{$d_{\perp}$}
  &\multicolumn{2}{c}{$d_{\parallel}$}
  &\multicolumn{2}{c}{$d_{-\perp}$}
  &\multicolumn{2}{c}{$d_{-\parallel}$}\\
  15 May 1997 &01:14:43 &01:15:23
  & 17 		& (2.6) 	&136 	& (21.2) 	&17 		& (2.6) 	&$\leq 140$ 	& (21.9)\\
  26 August 1998 &06:40:04 &06:40:27
  & 15 		& (2.3) 	&78 		& (12.2) 	&69 		& (10.1) 	&$\leq 26 $ 	& (4.0) \\
  11 February 2000 &23:32:55 &23:33:58
  & \nodata	& \nodata	& \nodata	& \nodata	&28 		& (4.3) 	&151 		& (32.6) \\
  \enddata

\end{deluxetable}

\clearpage


\end{document}